\documentclass[11pt]{article}
\usepackage{moriond,epsfig}
\usepackage{verbatim}
\usepackage{color}
\usepackage{rotate}
\usepackage{graphics}
\usepackage{epsf}
\usepackage{graphicx}
\usepackage{latexsym}
\usepackage{amscd}
\usepackage{amssymb}

\def\Journal#1#2#3#4{{#1} {\bf #2}, #3 (#4)}

\def\PRL{\em Phys. Rev. Lett.}
\def\PRD{{\em Phys. Rev.} D}
\def\PRC{{\em Phys. Rev.} C}

\def\be{\begin{equation}}
\def\ee{\end{equation}}
\def\bea{\begin{eqnarray}}
\def\eea{\end{eqnarray}}

\newcommand\eps\varepsilon

\begin{document}
\vspace*{4cm}
\title{THE COLOR DIPOLE APPROACH TO THE DRELL-YAN PROCESS IN $PA$ COLLISIONS}

\author{B.Z.~KOPELIOVICH$^{1,3}$, J.~RAUFEISEN$^2$ and A.V.~TARASOV$^{1,3}$ }

\address{
$^1$Max-Planck-Institut f\"ur Kernphysik, P.O. Box 103980, 69029 Heidelberg, Germany\\
$^2$Los Alamos National Laboratory, MS H846, Los Alamos, NM 87545, USA\\
$^3$Institut f\"ur Theoretische Physik, Universit\"at Regensburg, 93040 Regensburg,
Germany}

\maketitle\abstracts{In the target rest frame and at high energies, Drell-Yan (DY) dilepton
production looks like bremsstrahlung of massive photons, rather than
parton annihilation. 
The projectile quark is decomposed into a series of Fock states. 
Configurations with fixed transverse separations are interaction
eigenstates for $pp$ scattering. The DY cross section can then be expressed
in terms of the same color dipole cross section as DIS. This approach is
especially suitable to describe nuclear effects, since it allows to apply
Glauber multiple scattering theory.
We go beyond the Glauber eikonal approximation by taking into account
transitions between interaction eigenstates. 
We calculate nuclear shadowing at large Feynman-$x_F$ for DY in
proton-nucleus collisions, compare to existing data from E772 and make
predictions for RHIC. Nuclear effects on the transverse momentum
distribution are also investigated.
}

\section{Introduction}

\begin{figure}[ht]
  \scalebox{0.7}{\includegraphics*{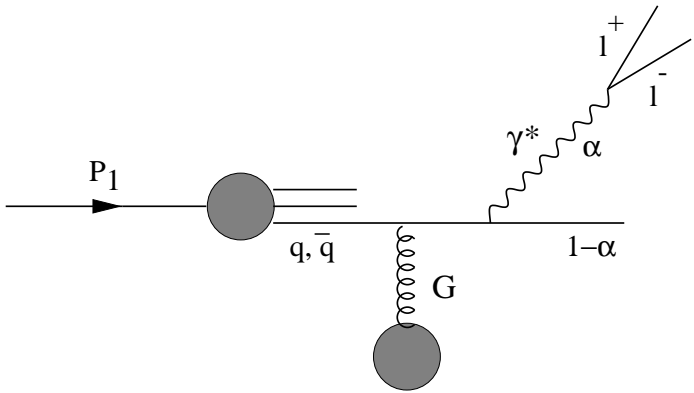}}\hfill
  \raise2.0cm\hbox{\parbox[b]{2.18in}{
    \caption{
      \label{bremsdy}
      A quark 
      (or an antiquark) inside the
      projectile hadron scatters off the target color field and radiates a
      massive photon, which subsequently decays into the lepton pair. The photon
      can also be radiated before the quark hits the target. Here, 
      {$\alpha$} is
      the {longitudinal momentum fraction} 
      of the quark carried by the photon.
    }
  }
}
\end{figure}

Although cross sections are Lorentz invariant, the partonic interpretation of the
microscopic process depends on the reference frame.  
As pointed out by one of the authors,
in the target rest frame DY
dilepton production should be treated as bremsstrahlung, rather than parton
annihilation \cite{boris} 
(see also \cite{bhq}). 
The space-time picture of the DY process in the target rest frame is
illustrated in fig.\ \ref{bremsdy}.  A quark (or an antiquark) from the projectile
hadron radiates a virtual photon on impact on the target. The radiation can occur before
or after the quark scatters off the target. Only the latter case is shown in
fig.~\ref{bremsdy}.

A salient feature of the rest frame picture of DY dilepton production is that at high
energies and in impact parameter space the DY cross section can be formulated in terms
of the same dipole cross section as low-$x_{Bj}$ DIS.  
%The transverse
%momentum distribution of DY dileptons can also be expressed in terms of this dipole cross
%section \cite{kst}.  

The color dipole approach to the DY process  provides 
a convenient alternative to the well known 
parton model, in
particular, it is especially appropriate to describe nuclear effects 
\cite{boris,kst}.

\section{DY dilepton production in $pp$ scattering}

The cross section for radiation of a virtual photon from a quark after
scattering on a proton, can be written in factorized light-cone form
\cite{boris,bhq,kst}, 
\be\label{dylctotal}
\frac{d\sigma(qp\to \gamma^*X)}{d\ln\alpha}
=\int d^2\rho\, |\Psi^{T,L}_{\gamma^* q}(\alpha,\rho)|^2
    \sigma_{q\bar q}(x_2,\alpha\rho),
\ee
similar to the case of DIS.
Here, $\sigma_{q\bar q}$ is the cross section \cite{mpb} for scattering  a
$q\bar q$-dipole off a proton which depends on the $q\bar q$ separation 
$\alpha\rho$,
where $\rho$ is  the photon-quark transverse separation and $\alpha$ 
is the fraction of 
the light-cone momentum of the initial quark taken away by the photon.
We use the standard notation for the kinematical variables,
$x_1-x_2=x_F$, $\tau=M^2/s=x_1x_2$, where $x_F$ is the
Feynman variable,
$s$ is the center of mass energy squared of the colliding protons and 
$M$ is the
dilepton mass. In (\ref{dylctotal}) $T$ stands for transverse and $L$
for longitudinal photons.

The physical interpretation of (\ref{dylctotal}) is similar to the DIS
case. The projectile quark is expanded in the
interaction eigenstates. We keep here only the first eigenstate,
\be
|q\rangle=\sqrt{Z_2}|q_{bare}\rangle+\Psi^{T,L}_{\gamma^* q}|q\gamma^*\rangle+\dots,
\ee
where $Z_2$ is the wavefunction renormalization constant for fermions.
In order to produce a new state the interaction must distinguish 
between the two Fock 
states, {\it i.e.} they have to interact differently. Since only the quarks
interact in both Fock components the difference arises from their relative displacement 
in the transverse plane.
 If $\rho$ is the transverse separation between the
quark and the photon, the $\gamma^*q$ fluctuation has a center of gravity in the
transverse plane which coincides with the impact parameter of the parent quark.
The transverse separation between the photon and the center of gravity is
$(1-\alpha)\rho$ and the distance between the quark and the center of gravity is
correspondingly $\alpha\rho$. Therefore, the argument of $\sigma_{q\bar q}$ is
$\alpha\rho$. More discussion can be found in \cite{last}.

The transverse momentum distribution of DY pairs
can also be expressed in terms of the dipole cross section \cite{kst}. 
The differential cross section is given by the 
Fourier integral
\bea\nonumber\label{dylcdiff}
\frac{d\sigma(qp\to \gamma^*X)}{d\ln\alpha d^2q_\perp}
&=&\frac{1}{(2\pi)^2}
\int d^2\rho_1d^2\rho_2\, \exp[{\rm i}\vec q_\perp\cdot(\vec\rho_1-\vec\rho_2)]
\Psi^*_{\gamma^* q}(\alpha,\vec\rho_1)\Psi_{\gamma^* q}(\alpha,\vec\rho_2)\\
&\times&
\frac{1}{2}
\left\{\sigma_{q\bar q}(x_2,\alpha\rho_1)
+\sigma_{q\bar q}(x_2,\alpha\rho_2)
-\sigma_{q\bar q}(x_2,\alpha(\vec\rho_1-\vec\rho_2))\right\}.
\eea
after integrating this expression over the transverse momentum
$q_\perp$ of the photon, one obviously recovers
(\ref{dylctotal}). 

\begin{figure}[ht]
\centerline{
  \scalebox{0.43}{\includegraphics{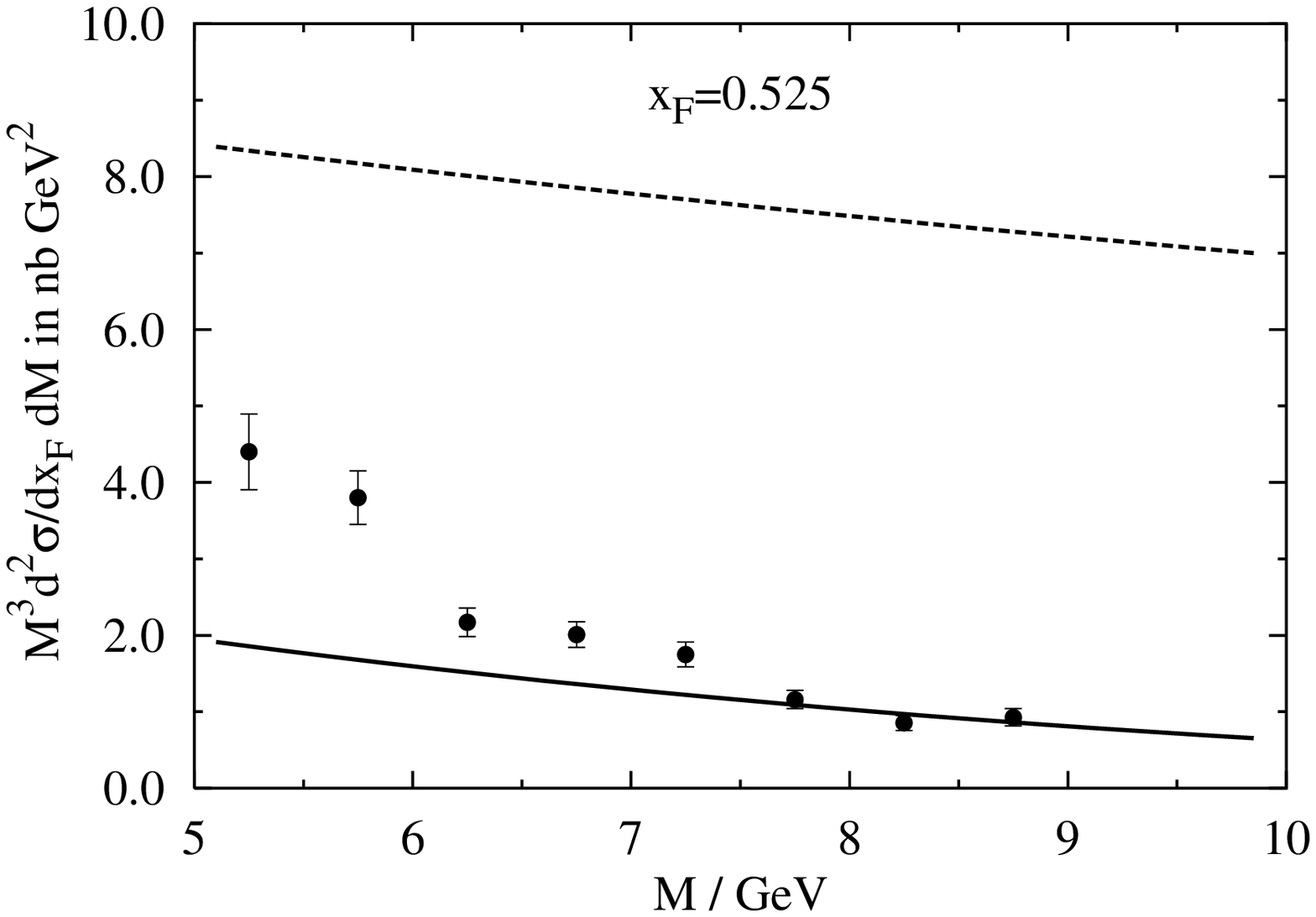}}
  \scalebox{0.43}{\includegraphics{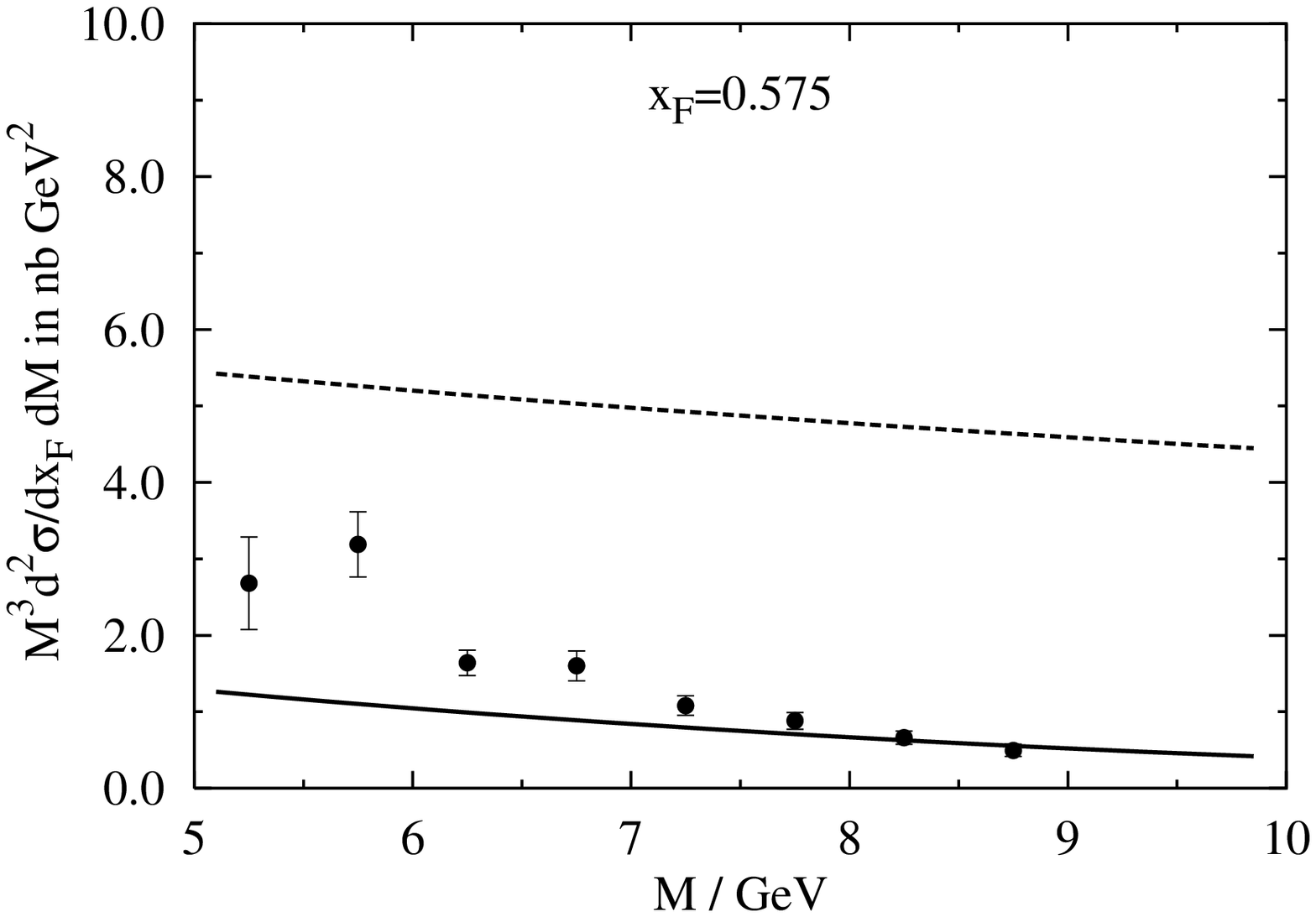}}
 }
\centerline{
  \scalebox{0.43}{\includegraphics{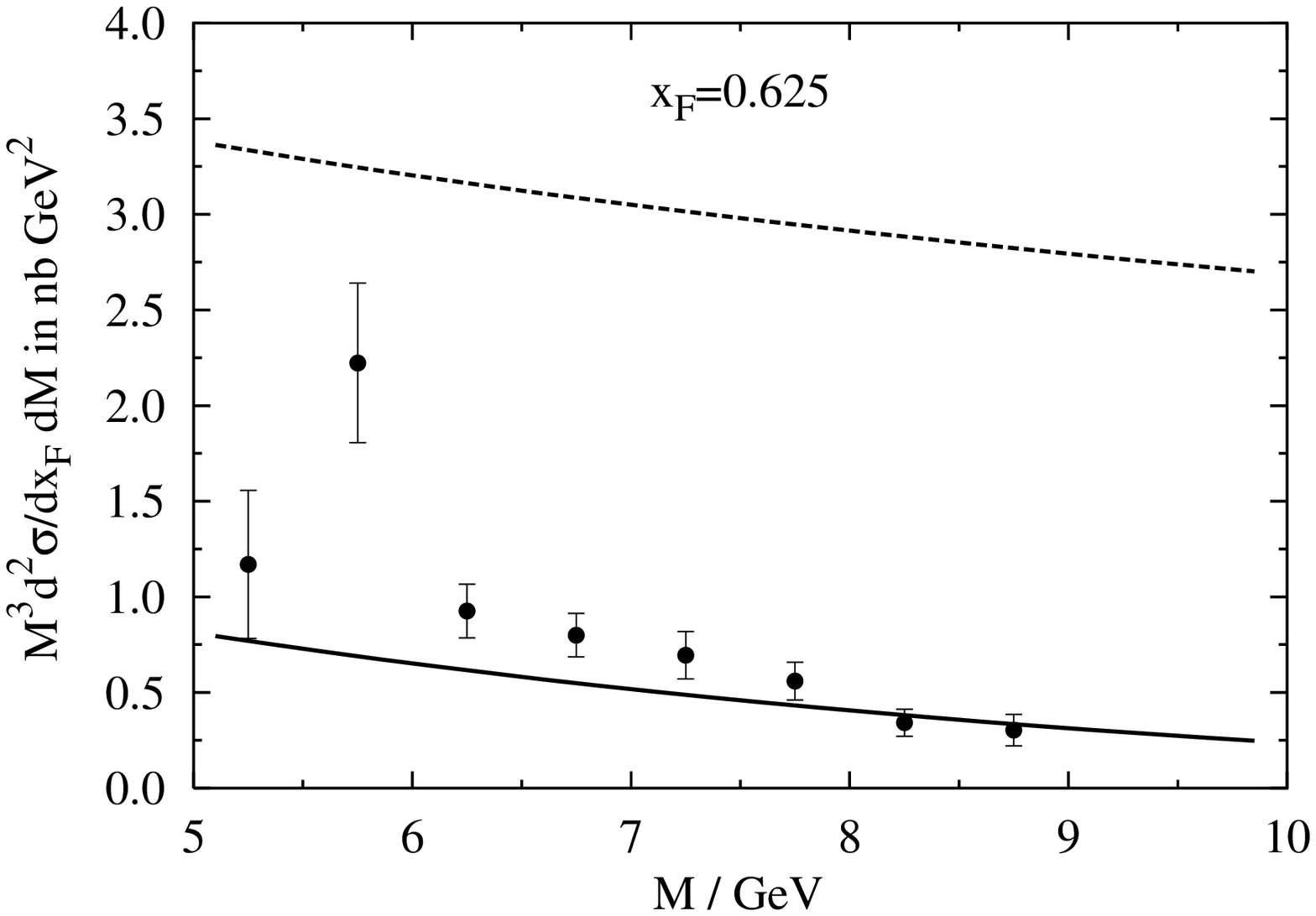}}
  \scalebox{0.43}{\includegraphics{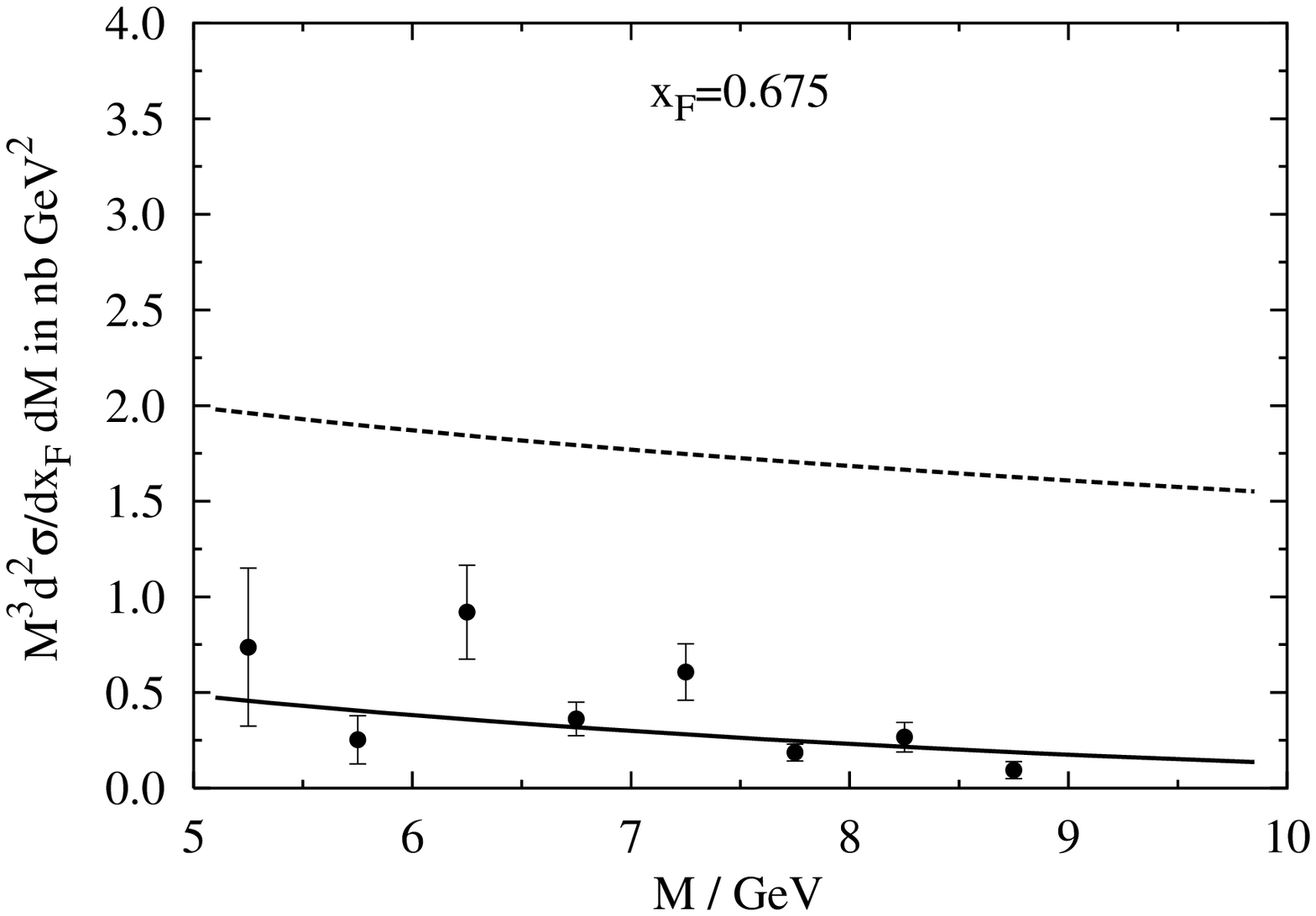}}
}
\caption{\label{dytotal}
  The points show the measured DY cross section in $p\,^2H$ scattering
  from E772. The curves are calculated  without any further
  fitting procedure.
  The solid curves are calculated at the same kinematics 
  as the data points (center
  of mass energy $\sqrt{s}=38.8$ GeV). The dashed curves are 
  calculated for RHIC
  energy, $\sqrt{s}=500$ GeV.
}
\end{figure}

The LC wavefunctions can be calculated in perturbation theory and are well known
\cite{bhq,last}.
The dipole cross section on the other hand is largely unknown.
Only at small distances $\rho$ it can be
expressed in terms of the gluon density. However, several 
successful parameterizations exist in the
literature, describing the entire function $\sigma_{q\bar q}(x,\rho)$, without
explicitly taking into account the QCD evolution of the gluon density.  
We use the
parameterization by Golec-Biernat and
W\"usthoff \cite{Wuesthoff1} for our calculations, fig.\ \ref{dytotal}.
This parameterization vanishes $\propto\rho^2$ at small distances, as
implied by color transparency \cite{mpb}
and levels off exponentially at large separations. 
The data in fig.\ \ref{dytotal} are quite well described 
without any $K$-factor, which does not appear in this approach since higher
order corrections are supposed to be parameterized in 
$\sigma_{q\bar q}(x_2,\rho)$.

\section{Proton-nucleus ($pA$) scattering}

Shadowing in DY is an interference phenomenon due to multiple scattering of the
projectile quark inside the nucleus. In the target rest frame, where DY dilepton production
is bremsstrahlung of massive photons, shadowing is the Landau-Pomeranchuk-Migdal (LPM)
effect. 
These interferences occur (fig.\ \ref{bremsdy}), because photons
radiated at different longitudinal coordinates $z_1$ and $z_2$ 
 are not
independent of each other. 
Thus, the amplitudes have to be added coherently. 
Destructive interferences can occur only if the longitudinal distance 
$z_2-z_1$
is smaller than the so called coherence length $l_c$, which is the time needed
to distinguish between a quark and a quark with a $\gamma^*$ nearby. 
It is given by the uncertainty relation,
\be\label{lc}
l_c=\frac{1}{\Delta
P^-}=\frac{1}{m_Nx_2}\,\frac{(1-\alpha)M^2}
{q_\perp^2+(1-\alpha)M^2+\alpha^2m_q^2}.
\ee
Here, $\Delta P^-$ is the light-cone energy denominator for the transition $q\to
q\gamma^*$ and $q_\perp$ is the relative transverse momentum of
the $\gamma^*q$ Fock state.
For $z_1-z_2>l_c$, the
radiations are independent of each other. 

An immediate consequence of this is that $l_c$ has to be
larger than the mean distance between two scattering centers in the nucleus
($\sim 2$ fm in the nuclear rest frame). Otherwise, the projectile quark could
not scatter twice within the coherence length and no shadowing would be observed.

We develop a Green function technique \cite{kst}, which allows one
to resum all multiple scattering terms,
similar to Glauber theory, and in addition treats the coherence length exactly.
The formalism is equivalent to the one proposed in \cite{Slava}
for the LPM effect in QED.
Our general expression for the nuclear DY cross section reads
\bea\nonumber
\frac{d\sigma(qA\to \gamma^*X)}{d\ln\alpha} & = &
A\,\frac{d\sigma(qp\to \gamma^*X)}{d\ln\alpha}- \frac{1}{2} {\rm Re}\int d^2b
\int_{-\infty}^{\infty} dz_1 \int_{z_1}^{\infty} dz_2
\int d^2\rho_1\int d^2\rho_2\,
\\ \nonumber
& \times &
\Bigl[\Psi_{\gamma^*q}\left(\alpha,
\rho_2\right)\Bigr]^*\,
\rho_A\left(b,z_2\right)\sigma_{q\bar{q}}\left(x_2,\alpha\rho_2\right)
{G\left(\vec \rho_2,z_2\,|\,\vec \rho_1,z_1\right)}\\
&\times &
\rho_A\left(b,z_1\right)\sigma_{q\bar{q}}\left(x_2,\alpha\rho_1\right)
\Psi_{{\gamma^*q}}\left(\alpha,\rho_1\right). 
\label{dyshadowing}
\eea

The first term is just $A$ times the single scattering cross section, where $A$ is the
nuclear mass number. The second term is the shadowing correction. The impact parameter is
$b$ and the nuclear density is $\rho_A$. The Green function $G$ describes, how the
bremsstrahlungs-amplitude at $z_1$ interferes with the amplitude at $z_2$. 

To make the meaning of Eq.\ \ref{dyshadowing} more clear, let us first
consider a limiting case for $G$. 
In the simplest case, the coherence length, Eq.\
\ref{lc}, is infinitely long 
and only the double scattering term is taken into account. Then 
$G\left(\vec \rho_2,z_2\,|\,\vec \rho_1,z_1\right)=\delta^{(2)}(\vec\rho_1-\vec\rho_2)$ and
one of the $\rho$ integrations can be performed. The $\delta$-function means that at very
high energy (infinite coherence length) the transverse size of the $\gamma^*q$ Fock-state
does not vary during propagation through the nucleus, it is frozen due to Lorentz time
dilatation. Furthermore, partonic configurations with fixed transverse separations in impact
parameter space were identified a long time ago 
\cite{mpb} in QCD as interaction eigenstates. 
This is the reason, why we work in coordinate space. Namely, in coordinate
space, all multiple scattering terms can be resummed and in the limit of 
infinite $l_c$ 
one obtains 
\be
G^{frozen}
\left(\vec\rho_2,z_2\,|\,\vec\rho_1,z_1\right)=\delta^{(2)}(\vec\rho_1-\vec\rho_2)
\exp\left(-\frac{\sigma_{q\bar q}(x_2,\rho_1)}{2}\int_{z_1}^{z_2}dz\rho_A(b,z)\right).
\ee 
The frozen approximation is identical to eikonalization of 
the dipole cross section in Eq.\
(\ref{dylctotal}). Thus, the impact parameter representation allows a very simple
generalization from a proton to a nuclear target, provided the coherence length is
infinitely long.

At Fermilab fixed-target energies ($\sqrt{s}=38.8$ GeV for E772), 
this last condition is not fulfilled and one has to take a finite $l_c$ into account. The
problem is however, that $l_c$, Eq.\ \ref{lc}, depends on the relative transverse momentum
$q_\perp$ of the $\gamma^*q$-fluctuation which is the conjugate variable to the size $\rho$
of this Fock-state and therefore completely undefined in $\rho$-representation. The quantum
mechanically correct way to treat the $q_\perp^2$ in Eq.\ \ref{lc} is to represent it by a
two-dimensional Laplacian $\Delta_\perp$ in $\rho$-space. The Green function which contains
the correct, finite coherence length and resums all multiple scattering terms fulfills a
two-dimensional Schr\"odinger equation with an imaginary potential,
\bea\nonumber
\left[{\rm i}\frac{\partial}{\partial z_2}
+\frac{\Delta_\perp\left(\rho_2\right)-\eta^2}
{2E_q\alpha\left(1-\alpha\right)}
+\frac{{\rm i}}{2}\rho_A\left(b,z_2\right)\,
\sigma_{q\bar{q}}\left(x_2,\alpha\rho_2\right)
\right]
{G\left(\vec \rho_2,z_2\,|\,\vec \rho_1,z_1\right)}\\ 
\qquad =
{\rm i}\delta\left(z_2-z_1\right)
\delta^{\left(2\right)}\left(\vec \rho_2-\vec \rho_1\right).
\label{sgl}
\eea
For details of the derivation, we refer to \cite{kst}.

The imaginary potential accounts for all higher order scattering terms. The Laplacian implies that
the Green function is no longer proportional to a $\delta$-function. This means the size
of the $\gamma^*q$ fluctuation is no longer constant during propagation through the nucleus.
One can say that an eigenstate of size $\rho_1$ evolves to an eigenstate of size
$\rho_2\neq\rho_1$, so transitions between eigenstates occur.

\begin{figure}[t]
  \scalebox{0.6}{\includegraphics{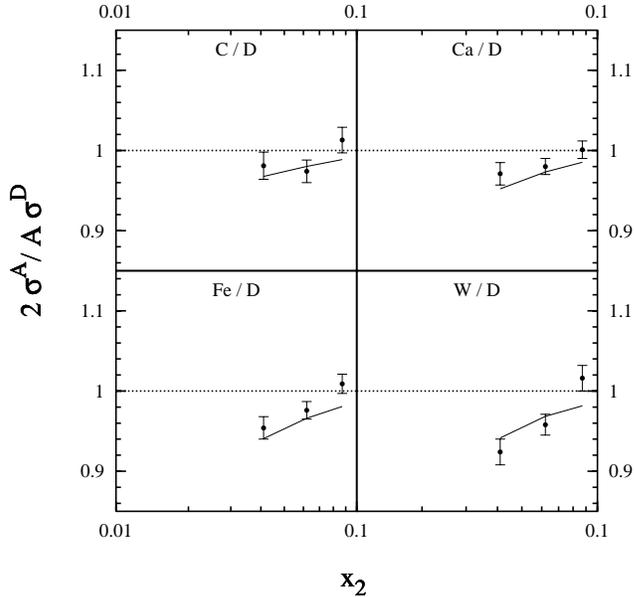}}\hfill
  \raise5cm\hbox{\parbox[b]{2.47in}{
   \caption{\label{e772}
   Comparison between calculations in the Green function technique and
      E772 data at 
center of mass energy {$\sqrt{s}=38.8$ GeV.} 
  for shadowing in DY. 
      The calculations are performed at the mean values of the lepton pair mass.
      From the
      left to the right, these values are $5$~GeV,
      $5.7$~GeV, and $6.5$~GeV.
	} 
  }
}
\end{figure}

Calculations \cite{thesis}
with Eqs.\ \ref{dyshadowing} and \ref{sgl} are compared to E772 data 
\cite{shadow} in fig.\
\ref{e772}. Note that the coherence length $l_c$ at E772 
energy becomes smaller than the nuclear radius. Shadowing vanishes as 
$x_2$ approaches $0.1$, because the coherence length becomes smaller 
than the mean internucleon separation.
It is therefore important 
to have a correct description of a finite $l_c$ in this energy range. 

\begin{figure}[ht]
  \scalebox{0.5}{\includegraphics*{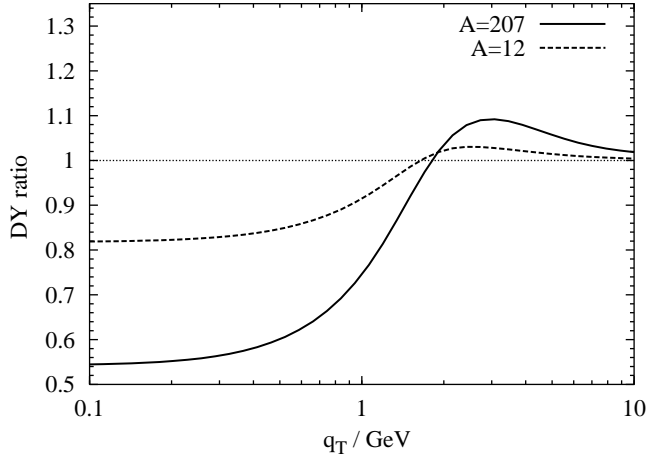}}\hfill
  \raise3.2cm\hbox{\parbox[b]{2.44in}{
   \caption{\label{ratio}
  Nuclear effects on the DY transverse momentum distribution at RHIC energy
  ($\sqrt{s}=200$ GeV) for carbon and lead. The figure 
  shows the DY cross section
  ratio $(d\sigma^A/dx_FdM^2d^2q_\perp) / (A d\sigma^p/dx_FdM^2d^2q_\perp)$ for
  dilepton mass $M=5$ GeV and Feynman $x_F=0.6$. Here, $q_\perp=q_T$ is the
      transverse momentum of the lepton pair.
	} 
  }
}
\end{figure}

Nuclear effects on the
$q_\perp$-differential cross section calculated at RHIC energy
are shown in fig.\ \ref{ratio}.
See \cite{kst} for details of the calculation. 
The differential cross section is suppressed at small
transverse momentum $q_\perp$ of the dilepton, 
where large values of $\rho$ dominate. This suppression
vanishes at intermediate $q_\perp\sim 2$ GeV. In this region, one even observes an
enhancement which reminds one of the Cronin effect. This enhancement is due to
multiple scattering of the quark inside the nucleus. A nuclear target 
provides a larger momentum transfer than a proton target and harder fluctuations
are freed, which leads to nuclear
broadening.
Note, that not the entire
suppression at low $q_\perp$ is due to shadowing. Some of the dileptons missing at low
$q_\perp$ reappear in this enhancement region. At very large 
transverse momentum nuclear
effects vanish.

\section{Summary}

We express the DY cross section in terms of the cross section $\sigma_{q\bar q}$ for
scattering a $q\bar q$ dipole off a proton. This is the same dipole cross section that
appears in DIS. We can reasonably well describe low $x_2$
DY data from $pp$ collisions \cite{dydata}, without any free parameters and without a $K$
factor. 

At very high energy, 
the dipole approach is easily extended to nuclear targets by eikonalization.
At lower fixed target energies (E772) the frozen approximation is no longer valid, because
the size of a Fock state  varies during propagation through the nucleus. Therefore,
transitions between interaction eigenstates (i.e.\ 
partonic configurations with fixed transverse
separations)  
occur.

We develop a Green function technique,
which takes variations of the transverse size into account and 
resums all multiple scattering
terms as well. Calculations with the Green function technique are in good agreement with DY
shadowing data from E772.
We have also calculated
nuclear effects in the transverse momentum distribution of DY pairs at RHIC
energy. The DY cross section is suppressed at low transverse momentum, but enhanced at
intermediate $q_\perp\sim 2$ GeV. Nuclear effects vanish at very large $q_\perp$.

\section*{Acknowledgments}

This work was partially supported by the 
Gesellschaft f\"ur Schwer\-ionenforschung, GSI, grants HD H\"UF T 
and GSI OR SCH,
by the European Network
{\em Hadronic Physics with Electromagnetic Probes,}
Contract No.~FMRX-CT96-0008, 
and by the U.S.~Department of Energy at Los Alamos
National Laboratory under Contract No.~W-7405-ENG-38. 

We are grateful to  J\"org H\"ufner, 
Mikkel Johnson and Andreas Sch\"afer for valuable discussions.

\end{document}